\documentclass[epj]{webofc}
\usepackage[utf8]{inputenc}
\usepackage[varg]{txfonts}   
\usepackage{booktabs}
\usepackage{xcolor}
\definecolor{darkred}{rgb}{0.4,0.0,0.0}
\definecolor{darkgreen}{rgb}{0.0,0.4,0.0}
\definecolor{darkblue}{rgb}{0.0,0.0,0.4}
\usepackage[bookmarks,linktocpage,colorlinks,
    linkcolor = darkred,
    urlcolor  = darkblue,
    citecolor = darkgreen]{hyperref}
\usepackage{subfigure}
\wocname{EPJ Web of Conferences}
\woctitle{Lattice2017}
\newcommand{\Xsl}[1]{\raise.15ex\hbox{/}\kern-.57em #1}

\begin{document}
%
\selectlanguage{english}
\title{%
Nucleon structure from 2+1-flavor domain-wall QCD
}
\author{%
\firstname{Shigemi} \lastname{Ohta}\inst{1,2,3}\fnsep\thanks{Speaker, \email{shigemi.ohta@kek.jp}}\ 
 for the RBC and UKQCD Collaborations
 }
\institute{%
Institute of Particle and Nuclear Studies, High-Energy Accelerator Research Organization (KEK), Tsukuba, Ibaraki 305-0801, Japan
\and
Department of Particle and Nuclear Physics, Sokendai Graduate University of Advanced Studies, Hayama, Kanagawa 240-0193, Japan
\and
RIKEN BNL Research Center, Brookhaven National Laboratory, Upton, NY 11973, USA
}
\abstract{%
Nucleon-structure calculations of isovector vector- and axialvector-current form factors, transversity and scalar charge, and quark momentum and helicity fractions are reported from two recent 2+1-flavor dynamical domain-wall fermions lattice-QCD ensembles generated jointly by the RIKEN-BNL-Columbia and UKQCD Collaborations with Iwasaki \(\times\) dislocation-suppressing-determinatn-ratio gauge action at inverse lattice spacing of 1.378(7) GeV and pion mass values of 249.4(3) and 172.3(3) MeV.

\vspace{-119mm}\parbox{0.95\textwidth}{\flushright\rm KEK-TH-2006, RBRC-1253}\vspace{112mm}
}
\maketitle
\section{Introduction}\label{intro}

The RIKEN-BNL-Columbia (RBC) collaboration 
have been investigating nucleon structure using the domain-wall fermions (DWF) quarks on a sequence of quenched \cite{Sasaki:2003jh,Orginos:2005uy} and 2- \cite{Lin:2008uz} and 2+1-flavor \cite{Yamazaki:2008py,Yamazaki:2009zq,Aoki:2010xg} dynamical DWF ensembles at various mass values \cite{Blum:2000kn,Aoki:2004ht,Allton:2008pn,Aoki:2010dy,Arthur:2012yc,Blum:2014tka}.
As is well known, the DWF scheme allows to maintain continuum-like flavor and chiral symmetries on the lattice, and helps to simplify non-perturbative renormalizations \cite{Martinelli:1994ty,Blum:2001sr,Aoki:2007xm,Sturm:2009fk}.

In our earlier works calculated with degenerate up- and down-quark mass set at considerably heavier than physical values \cite{Yamazaki:2008py,Yamazaki:2009zq,Aoki:2010xg}, we observed the vector-current form factors behaving reasonably well in trending toward experiments: both Dirac and Pauli mean-squared charge radii and the isovector anomalous magnetic moment appeared to linearly depend on the pion mass squared.
The radii extrapolated to about 25 \% undershooting the experimental value at the physical mass.
It is interesting if the present calculations with considerably lighter mass improve these.

The axialvector-current form factors were found more problematic.
We saw significant deficit in calculated axial charge, \(g_A\), and form factors in general appear more susceptible to finite-size effect than the vector-current ones \cite{Yamazaki:2008py,Yamazaki:2009zq}.
These observations now have been confirmed by several other major collaborations \cite{Dragos:2016rtx,Bhattacharya:2016zcn,Liang:2016fgy,thiscontribTsukamoto} using different actions but with similar lattice spacings and quark masses.
Especially important for calculations with Wilson-fermion quarks \cite{Dragos:2016rtx,Bhattacharya:2016zcn,thiscontribTsukamoto} is to remove the \(O(a)\) systematic errors at the linear order in the lattice spacing, \(a\) \cite{Liang:2016fgy,thiscontribTsukamoto}. 

For low moments of the structure functions \cite{Aoki:2010xg}, we found both quark momentum and helicity fractions show interesting trend toward the experiment at the lightest mass we calculated, away from stubbornly flat mass-independence away from the experimental value at higher mass.
Similarly the transversity showed interesting downward departure at the lightest mass away from the flat higher-mass values.
If this trending continues in our present calculations at considerably lighter mass toward the experiment is obviously an interesting question.
Transversity and ``scalar charge'' are relevant in search for new physics beyond the standard model such as neutron electric dipole moment \cite{Bhattacharya:2016zcn,Yoon:2016jzj}.
\begin{figure}[t]
\includegraphics[width=\columnwidth,clip]{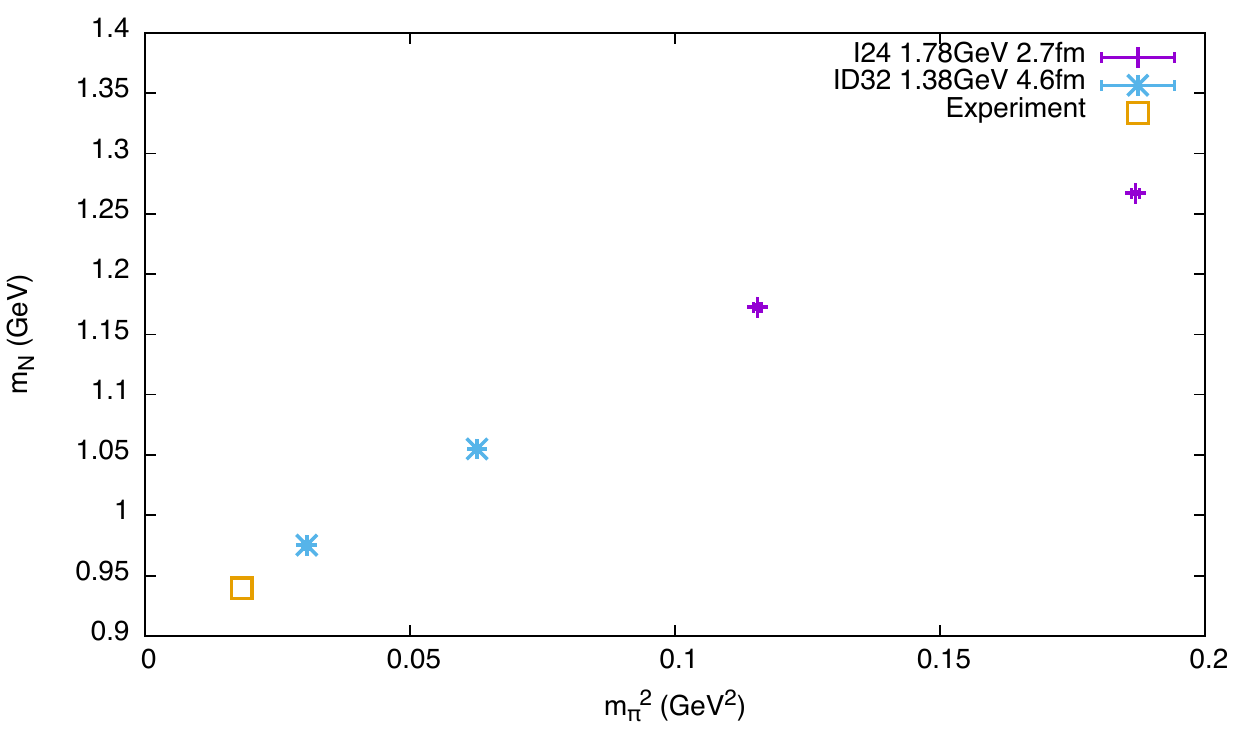}
\caption{
\label{fig:mpi2mN}
Estimated nucleon mass plotted against estimated pion mass squared of the present ensembles (ID32, cyan) and the two lightest of ref.\ \cite{Yamazaki:2009zq,Aoki:2010xg} (I24, magenta) with the new and more accurate estimates for the inverse lattice spacing \cite{Blum:2014tka}. 
The present results linearly extrapolate to the experiment (\(\Box\)) within the statistical error.
}
\end{figure}

In this talk we report nucleon isovector form factors of the vector and axialvector currents
and two lowest moments of isovector structure functions, namely quark momentum, \(\langle x \rangle_{u-d}\), and helicity, \(\langle x \rangle_{\Delta u - \Delta d}\), fractions, and isovector transversity, \(\langle 1 \rangle_{\delta u - \delta d}\), and isovector ``scalar charge,'' \(g_S^{u-d}\), calculated using two recent 2+1-flavor dynamical DWF lattice-QCD ensembles generated jointly by the RBC and UKQCD Collaborations with Iwasaki \(\times\) dislocation-suppressing-determinatn-ratio (DSDR) gauge action  on \(32^3\times64\) four-dimensional volume.
The inverse-squared gauge coupling of \(\beta=1.75\) resulted in the lattice momentum cut off of \(a^{-1}=1.378(7)\) GeV \cite{Arthur:2012yc,Blum:2014tka}.
Note the inverse lattice spacing has been slightly revised from the original \cite{Aoki:2010dy,Arthur:2012yc} by the global chiral and continuum fits in conjunction with new physical-mass ensemble sets \cite{Blum:2014tka} with M\"obius DWF quarks.
The strange-quark mass is set at 0.045, and degenerate up- and down-quark mass of 0.0042 and 0.001 in lattice units. 
Thus the heavier of the two ensembles corresponds to the pion mass, \(m_\pi\), of 249.4(3) MeV and spatial lattice extent \(L\) of \(m_\pi L = 5.79(6)\), and the lighter to 172.3(3) MeV and 4.00(6), respectively.
Our measurement calculations are made conventionally for the heavier ensemble for 165 configurations between the hybrid moleular-dynamics (MD) trajectory 608 and 1920 with 8-trajectory interval, each with seven source positions in time, and for the lighter with AMA for 39 configurations between 748 and 1420 with 16-trajectory interval for the lighter ensemble with 112 sloppy measurements each.

Some preliminary analyses of these nucleon-structure observables had been reported at recent Lattice conferences \cite{Lin:2014saa,Ohta:2013qda,Ohta:2014rfa,Ohta:2015aos,Abramczyk:2016ziv}.
In addition the LHP collaboration also calculated some nucleon structure \cite{Syritsyn:2009mx} using a RBC+UKQCD 2+1-flavor dyamical DWF ensemble \cite{Allton:2008pn}.

\section{Nucleon mass}

The nucleon mass for the present ensembles are 1.0550(20) and 0.9752(11) GeV respectively (see Fig.\ \ref{fig:mpi2mN}.)
Though not yet taken to the continuum limit, the estimates linearly extrapolate to the experiment.

\section{Form factors}

As had been reported in the earlier Lattice conferences, the vector-current Dirac and Pauli form factors, \(F_1\) and \(F_2\), behave reasonably well numerically \cite{Lin:2014saa,Ohta:2013qda,Ohta:2014rfa,Ohta:2015aos,Abramczyk:2016ziv} and allow parametrization in the conventional dipole function form \cite{Chambers:1956zz,Hofstadter:1961zz},  
\(
\propto (1+q^2/M_i^2)^{-2},
\)
where \(M_i (i=1, 2)\) are the dipole masses.
Preliminary results of such conventional dipole fits are presented, as in Tab.~\ref{tab:dipolefit}.
\begin{table}[t]
\caption{
\label{tab:dipolefit}
Dipole-fits for the vector, Dirac and Pauli, and axialvector form factors.
}
\begin{center}
\begin{tabular}{llllllll}
\hline
\multicolumn{1}{c}{\(m_\pi\) (MeV)} &
\multicolumn{1}{c}{\(M_{1}\) (GeV)} &
\multicolumn{1}{c}{\(r_{1}\) (fm)} &
\multicolumn{1}{c}{\(F_2(0)\)} &
\multicolumn{1}{c}{\(M_{2}\) (GeV)} &
\multicolumn{1}{c}{\(r_{2}\) (fm)} &
\multicolumn{1}{c}{\(M_{A}\) (GeV)} &
\multicolumn{1}{c}{\(r_{A}\) (fm)}\\
\hline\hline
172 & 1.09(8) & 0.63(5) & 3.3(2) & 0.81(6)	 & 0.84(6) & 1.23(14) & 0.55(6)\\
249 & 1.1(2) & 0.62(11) & 3.2(3) & 0.89(12)	& 0.77(9)& 1.26(15) & 0.54(6)\\
\hline
\end{tabular}
\end{center}
\end{table}

In common with the vector current, we use the local-current definition for the axialvector current.
Becuase the two local currents are connected by the chiral rotation, they share the common renormalization, \(Z_A=Z_V\), that relates them with the corresponding conserved global currents, up to \(O(a^2)\) descretization.
This is an advantage of the DWF scheme.
Thus for the axial charge, \(g_A\), it is better to look at its ratio, \(g_A/g_V\), to the vector charge, for precision.
\begin{figure}[b]
\includegraphics[width=0.48\columnwidth]{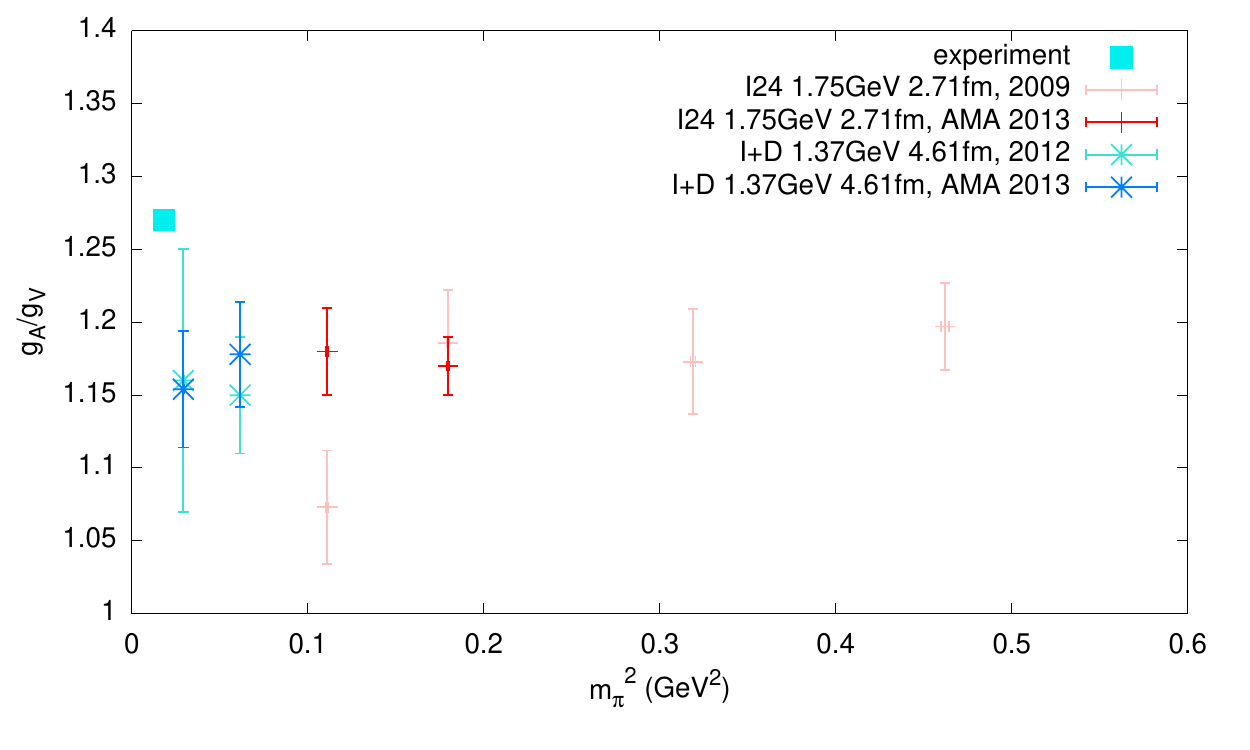}
\includegraphics[width=0.48\columnwidth]{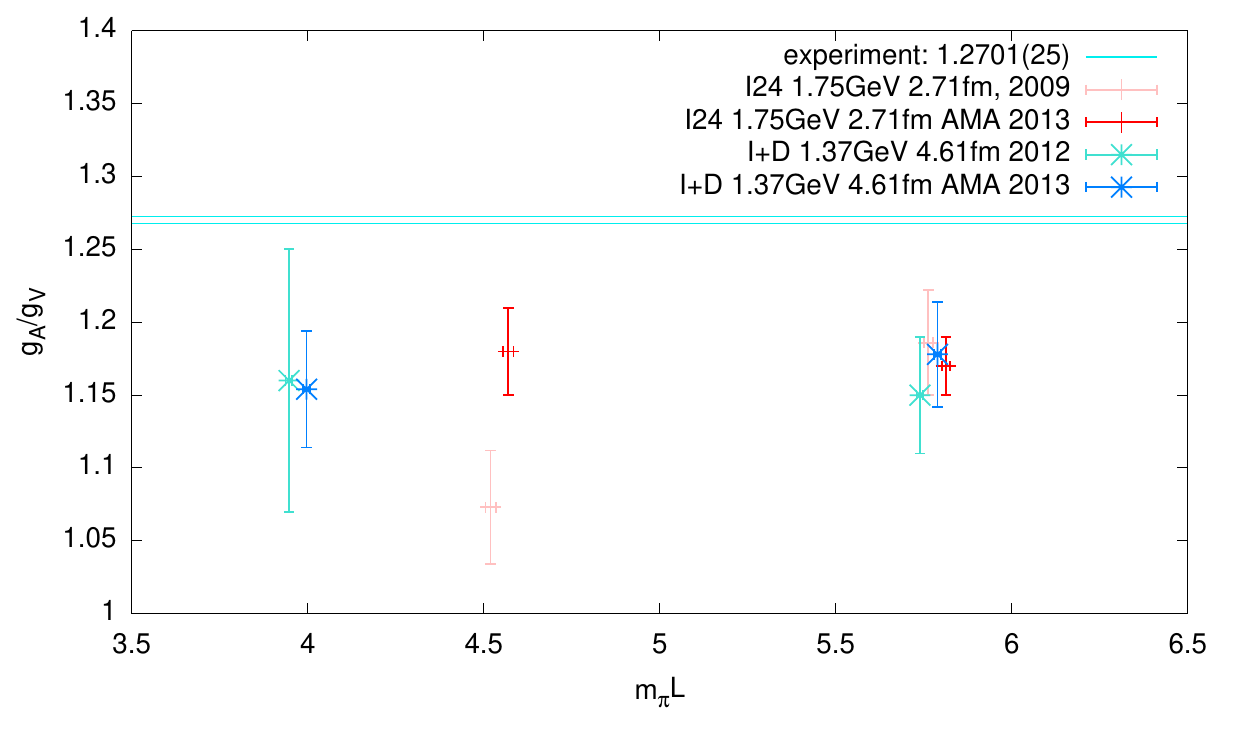}
\caption{\label{fig:AVmpi}
Dependence of the ratio, \(g_A/g_V\), of isovector axial charge, \(g_A\), and vector charge, \(g_V\), calculated with recent RBC+UKQCD 2+1-flavor dynamical DWF ensembles, on the pion mass squared (left) and chiral finite-size parameter, \(m_\pi L\), (right) in this talk and our earlier reports.
The experimental value quoted here is 1.2701(25) from PDG 2014 rather than the latest 1.2723(23) since PDG 2015 \cite{Patrignani:2016xqp}.
}
\end{figure}
The calculated values of the ratio, \(g_A/g_V\), underestimates the experimental value of 1.2723(23) \cite{Patrignani:2016xqp} by about 10 \% and do not depend much on the pion mass, \(m_\pi\), as shown in Fig.\ \ref{fig:AVmpi}, in the range from about 432 MeV down to 172 MeV,  from four recent RBC+UKQCD 2+1-flavor dynamical DWF ensembles \cite{Allton:2008pn,Aoki:2010dy,Arthur:2012yc}.

We are obviously suffering from some systematics that make our calculations undershoot the experimental value.

\begin{figure}[t]
\begin{center}
\includegraphics[width=0.48\textwidth]{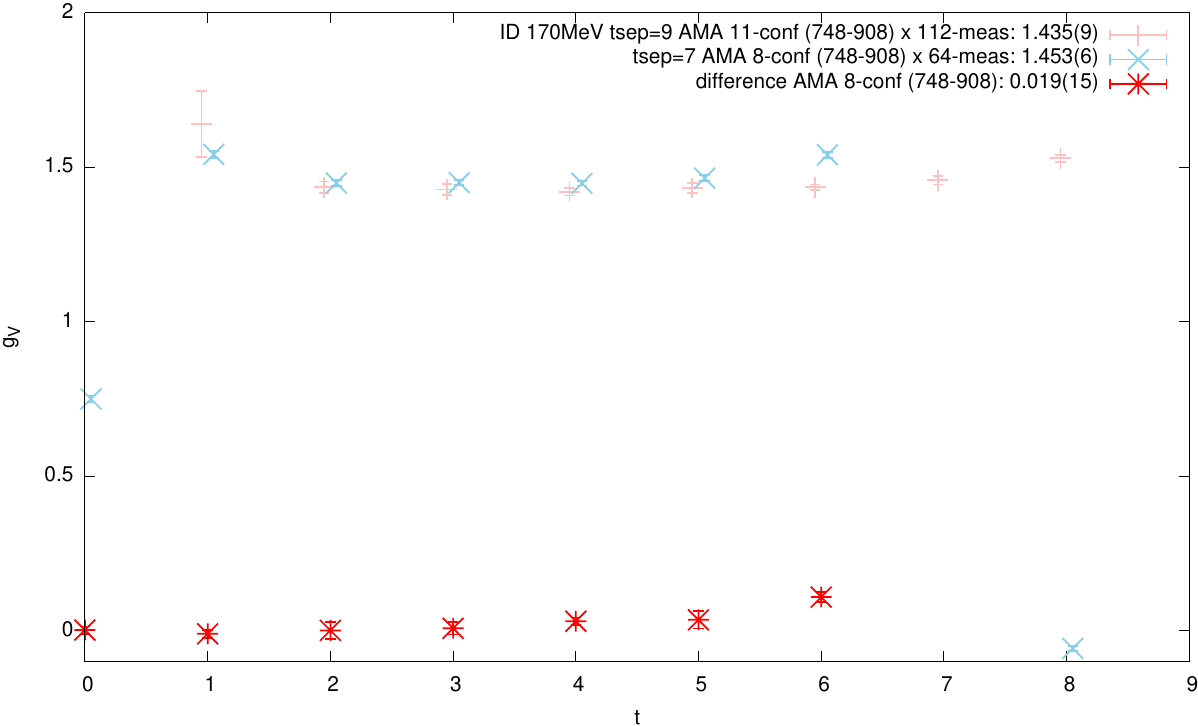}
\includegraphics[width=0.48\textwidth]{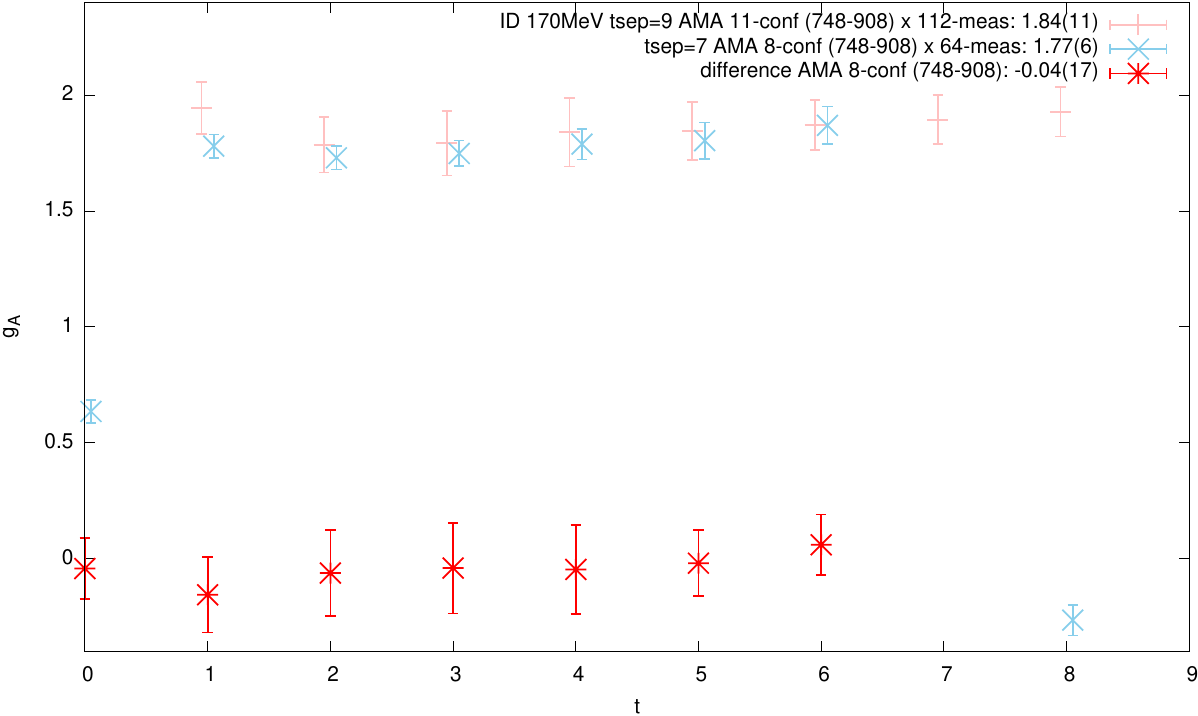}\\
\end{center}
\begin{center}
\includegraphics[width=0.48\textwidth]{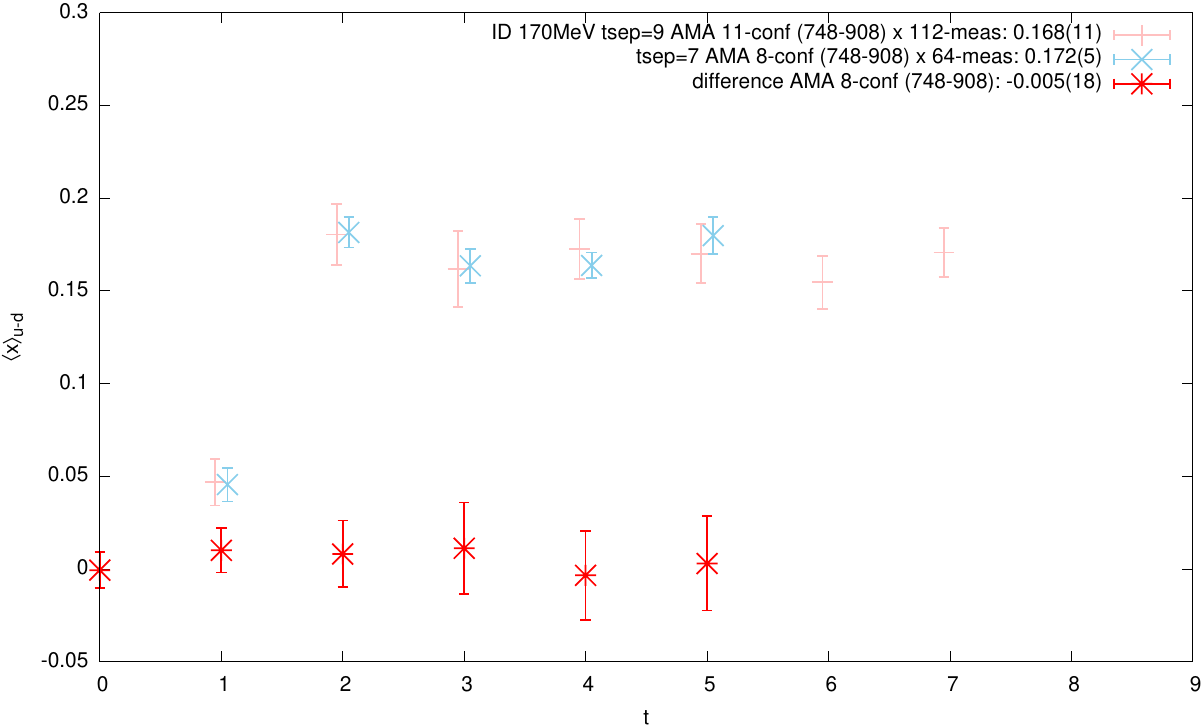}
\includegraphics[width=0.48\textwidth]{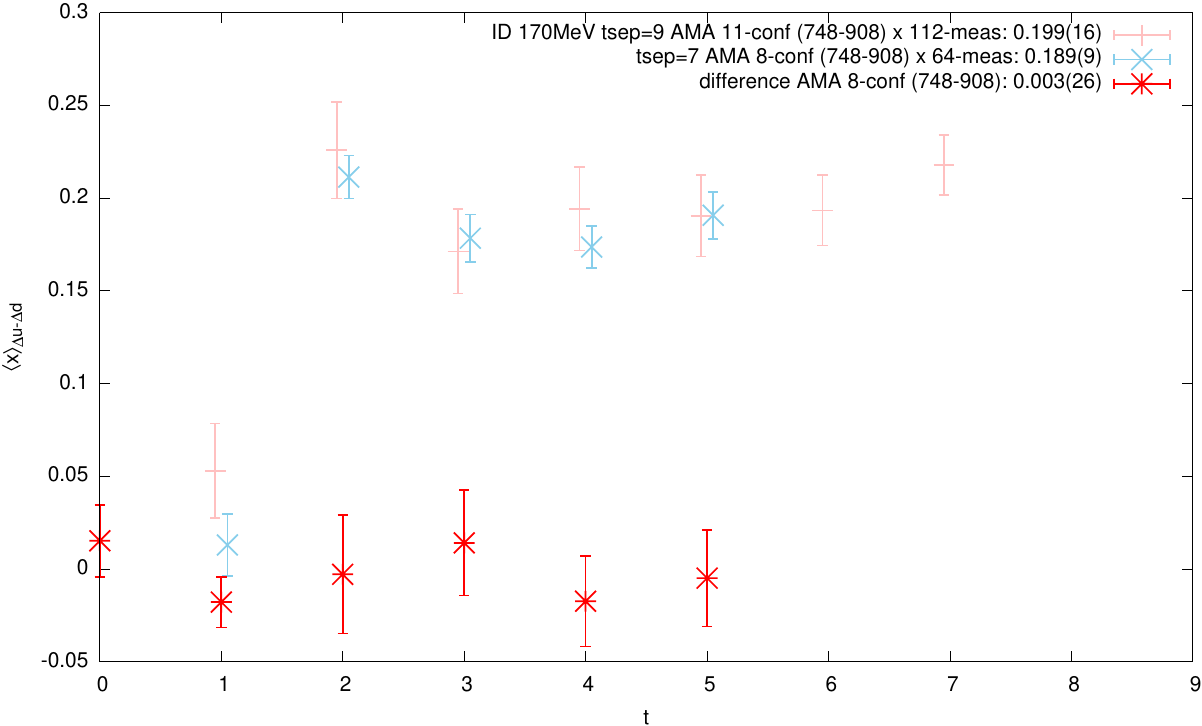}\\
\end{center}
\caption{\label{fig:noexcited}
No excited-state contamination is present in any of our nucleon isovector observables in 172-MeV ensemble: the results with source-sink separations of seven and nine never differ with each other.
}
\end{figure}

We know very well what systematics it is not: excited-state contamination.
When we set the lattice action with the coupling and mass, we set the excited-energy spectrum, \(\Delta_n = E_n-E_0\), for all the excited states \(n \ge 1\), as well as the ground-state energy \(E_0 = m_N\) of a nucleon.
Then when we decided the source/sink smearing, we decided the relative amplitude of the relevant excited states, \(|0\rangle + a_1|1\rangle + ...\) in what we are looking at.
These are not dependent on what observable we are looking at, and so can be be quantified from our measurements, especially when we can vary the source-sink separations, as the relative amplitude of any excited state damps faster than the ground, \(\propto a_n \exp(-\Delta_n t)\).
Of course some observable such as the conserved charge, \(g_V\), are diagonal, \(\langle n|g_V|0\rangle = 0\) for all \(n>0\), and do not suffer from any excited-state contamination.
Indeed the axial charge, \(g_A\), being a partially conserved charge, would be rather insensitive, \(\langle n|g_V|0\rangle \sim 0\) for all \(n>0\), and it would be an achievement to detect such contamination there.
We never achieved this.
We have however detected such contamination in other quantities such as quark momentum fraction \cite{Lin:2008uz} by comparing two different source-sink separations.
We had improved this with the current ensembles \cite{Lin:2014saa,Ohta:2013qda,Ohta:2014rfa,Ohta:2015aos,Abramczyk:2016ziv}, and demonstrated the absence of such excited-state contamination in all the observables we are reporting here as in Fig.\ \ref{fig:noexcited}.
The difference between the results from source-sink separation of seven and nine is always consistent with 0.
The contamination, \(a_n \langle n | O | 0 \rangle\), are either all negligible, or magically canceling each other  for all the observables we are looking at.

\begin{figure}[tb]
\includegraphics[width=0.48\columnwidth]{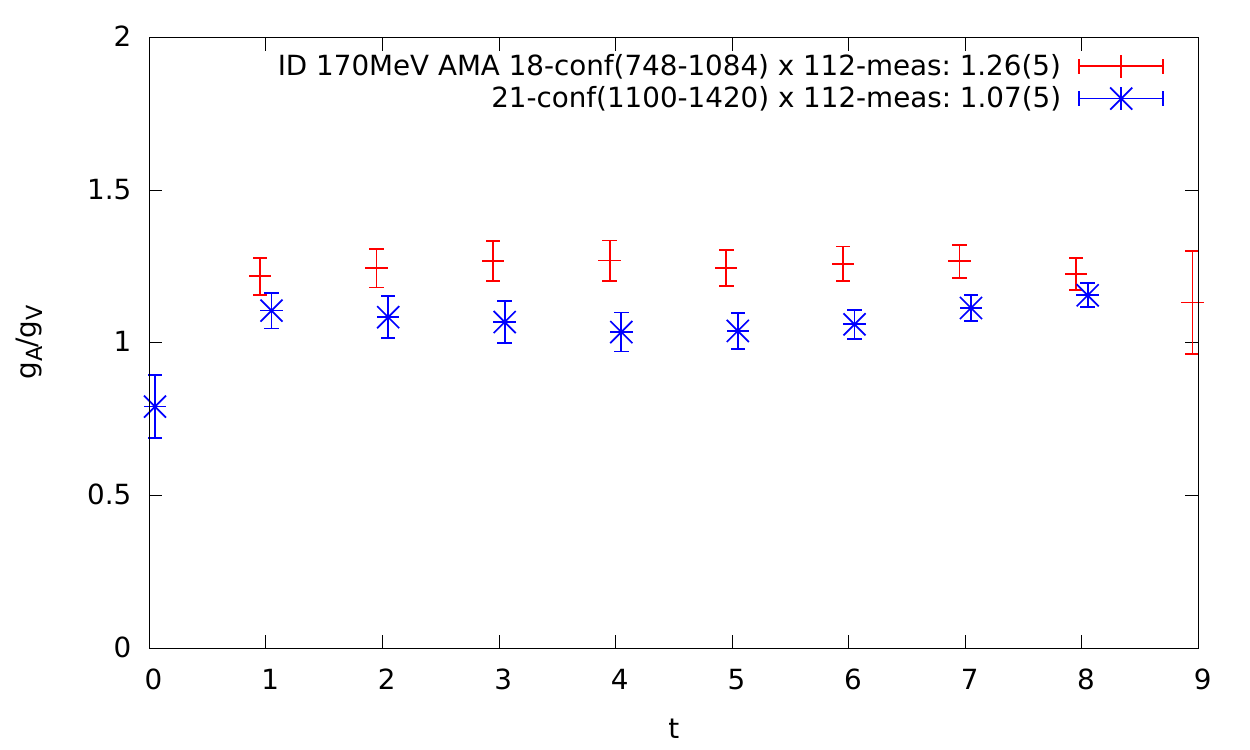}
\includegraphics[width=0.48\columnwidth]{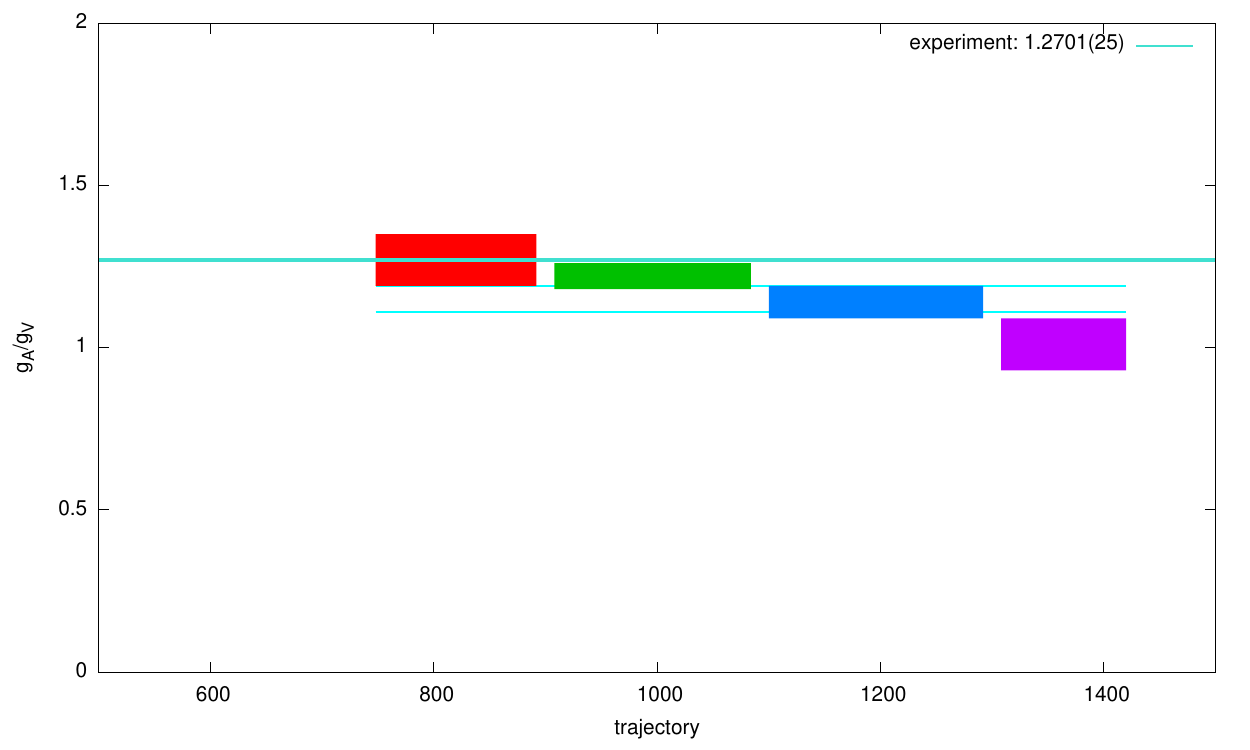}
\caption{\label{fig:LongAC}
Left: plateaux of the ratio, \(g_A/g_V\), for the first (trajectory from 748 to 1084, red) and the second (1100 to 1420, blue) halves, respectively, fitted in the range from 2 to 7 lattice units, the values of 1.26(5) for the first and 1.07(5) for the second are almost four standard deviations away from each other.
Right: Quarter-wise average along the hybrid MD time, from 748 to 892, 908 to 1084, 1100 to 1292, and 1308 to 1420: the values seem to drift monotonically from what is consistent with the experiment as of 2014 PDG of 1.2701(25) or the latest 1.2723(23) since 2015 PDG in the first quarter to a value around unity in the last quarter.
}
\end{figure}

In contrast to this absence of excited-state contamination, we see possible signs of inefficient sampling:
First we observe an unusually long-range autocorrelation when we divide the lightest ensemble at \(m_\pi\) = 172 MeV into two halves, earlier and later, in hybrid MD time, as in  Fig.~\ref{fig:LongAC}.
Indeed when we further divide into four consecutive quarters in MD time, the axial charge start at a value consistent with experiment but monotonically decrease to a value around unity.
We also note that no such under-sampling is seen in most other isovector observables we have looked at, the vector charge, \(g_V\), quark momentum fraction, \(\langle x \rangle_{u-d}\) and quark helicity fraction, \(\langle x \rangle_{\Delta u-\Delta d}\), and that blocked-jackknife analyses with block size of 2 showed strong correlation of two successive gauge configurations for \(g_A\) and \(g_A/g_V\), and possibly for transversity \(\langle 1 \rangle_{\delta u - \delta d}\), but not for the other.

A similar but weaker sign of unusually long-range autocorrelation is seen in an earlier ensemble \cite{Allton:2008pn} at \(m_\pi\) = 340 MeV when we divide it into four consecutive quarters in hybrid MD time.
However no such sign of under-sampling is seen in the other ensembles including the heavier of the present ones at \(m_\pi\) = 249 MeV.
In other words, the strongest sign of under-sampling is seen at the smallest of finite-size scaling parameter, \(m_\pi L \sim 4.00(6)\), another weaker sign seen at the second smallest but not the second lightest at \(m_\pi L \sim 4.569(15)\), and not seen in larger values at \(> m_\pi L \sim 5.79(6)\).
This of course does not prove the problem is caused by the finite lattice spatial volume, but suggests so.

That there is a long-range autocorrelation in this observable is corroborated by blocked-jack-knife analysis with block sizes of 2, 3, and 4: the statistical error of the axial charge keeps growing to at least beyond block size of 3 while those for the other observables do not grow at all except perhaps for transversity which nonetheless stops growing earlier.

If an observable appears long-range autocorrelated, it would be interesting to look at its correlation with the topology of the gauge configurations.
We explored this possibility by plotting jackknife samples against topological charge and did not find correlation.

We can also look at if our low-mode deflation affected this, though the available information is limited to about half of the configurations of what we are presenting from the 172-MeV ensemble
Albeit with this limitation we do not find any correlation either. 

It may be also instructive to remember earlier phenomenological analyses such as by the MIT bag model that estimates \(g_A/g_V = 1.09\) without pion \cite{Chodos:1974pn}, and another by the Skyrmion model that gives only conditionally convergent result of 0.61, that is strongly dependent on pion geometry \cite{Adkins:1983ya}.
To explore such spatial dependence arising from pion geometry, we divided the AMA samples into two spatial halves such as \(0\le x < L/2\) and\(L/2 \le x < L\) for each of the three spatial directions in order to check if there is any uneven spatial distribution
We find the calculation fluctuates in space.
Larger spatial volume are likely to stabilize the calculation better.

To summarize: we explored possible causes of about 10-\% deficit in lattice-QCD calculations of nucleon isovector axial and vector charge ratio, \(g_A/g_V\), in comparison with the experiment of 1.2723(23).
As we reported at Lattice 2013, an unusually long-range autocorrelation was seen in our lightest, 172-MeV, 2+1-flavor dynamical DWF ensemble.
It is also present in the 340-MeV ensemble with the second smallest \(m_\pi L\), but is not present in the other two, 249-MeV and 432-MeV ensembles with larger \(m_\pi L\).
No other isovector observable shows such a problem, except perhaps transversity where the effect is weak at worst.
No correlation is seen with gauge-field topology nor low-mode deflation.
In contrast the axial-charge calculation appears to fluctuate spatially along the course of molecular dynamics evolution.

We also report the isovector axialvector and induced pseudoscalar form factors.
The axialvector form factor can be parameterized in much the same way in the conventional dipole form as the vector-current ones, and are tabulated along with them in Tab.\ \ref{tab:dipolefit}.
The induced pseudoscalar form factor exhibits a strong pion-pole behavior, as is expected from PCAC current algebra with pion-pole dominance for \(m_{\pi}\approx 0\).
Our preliminary estimates for the pseudoscalar coupling is \(g_P = 7.1(1.1)\) and 5.11(15).

\section{Transversity and scalar charge}

\begin{figure}[t]
\includegraphics[width=0.48\textwidth]{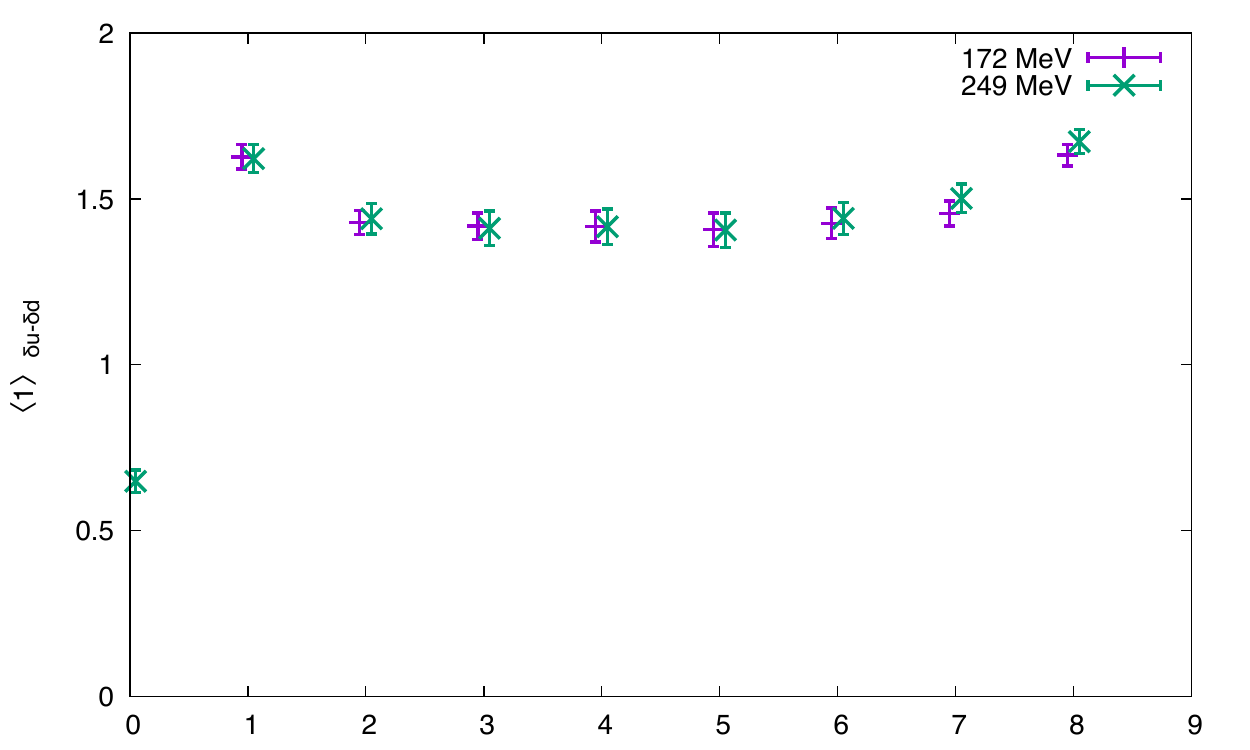}
\includegraphics[width=0.48\textwidth]{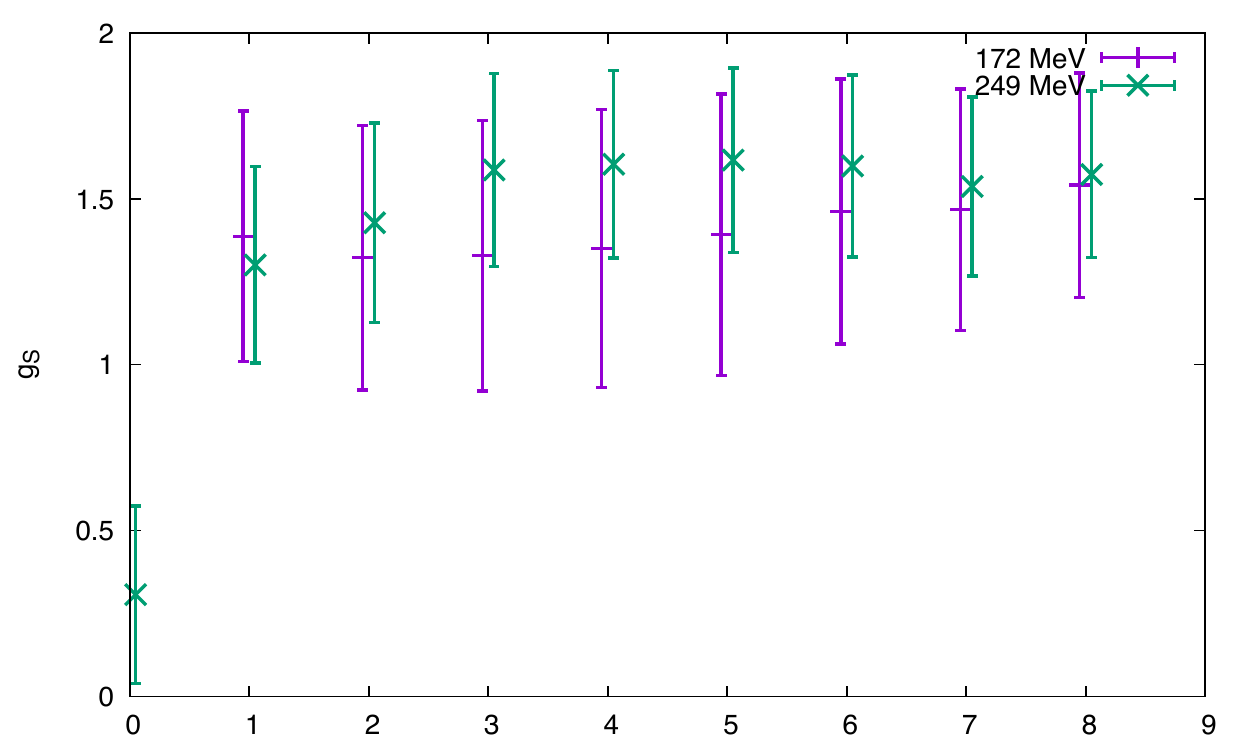}
\caption{\label{fig:1qgS}
Isovector transversity (left) and scalar charge (right) signals.  There is no mass dependence.
}
\end{figure}

The transversity signals are very clean and the ``scalar charge'' plateaux are well defined albeit with larger statistical errors (see Fig.\ \ref{fig:1qgS}.)
\begin{table}[b]
\begin{center}
\begin{tabular}{lll}
\hline
\multicolumn{1}{c}{\(m_\pi\) [MeV]} &
\multicolumn{1}{c}{\(\langle 1 \rangle_{\delta u-\delta d}\)} &
\multicolumn{1}{c}{\(g_S\)}\\
\hline\hline
172& \(1.42(4) \times 0.73(3) = 1.05(5)\) & \(1.4(4) \times 0.62(3) = 0.9(3)\) \\
249& \(1.42(5) \times 0.73(3) = 1.05(5)\) & \(1.6(3) \times 0.62(3) = 1.0(2)\) \\
\hline
\end{tabular}
\end{center}
\caption{\label{tab:TS}
Renormalized isovector transverstiy, \(\langle 1 \rangle_{\delta u - \delta d}\), and scalar charge, \(g_S\).
}
\end{table}
Neither shows mass dependence.
As had been reported, the isovector tranversity shows weaker but still relevant signs of long-lasting autocorrelation similar to that of the axial charge in the lighter, 172-MeV, ensemble \cite{Ohta:2014rfa}.
Yet the agreement with the heavier ensemble where there is no such autocorrelation reassures this is less problematic here in the transversity than in the axial charge.
With non-perturbative renormalizations of
\(
Z_S({\rm RI/SMOM}, \mu={\rm 2.0 GeV})=0.619(08)_{\rm stat}(24)_{\rm syst},
\)
and
\(
Z_T({\rm RI/SMOM}, \mu={\rm 2.0 GeV})=0.731(08)_{\rm stat}(29)_{\rm syst},
\)
we obtain our estimates as in Tab.\ \ref{tab:TS}.
The transversity errors are dominated by a scheme-dependence systematics in non-perturbative renormalization, at about five percent, due mainly from the relatively low lattice cut off.
The scalar errors are still dominated by statistical noise.  

\section{Low moments of Structure-functions}

\begin{figure}[t]
\includegraphics[width=0.48\textwidth,clip]{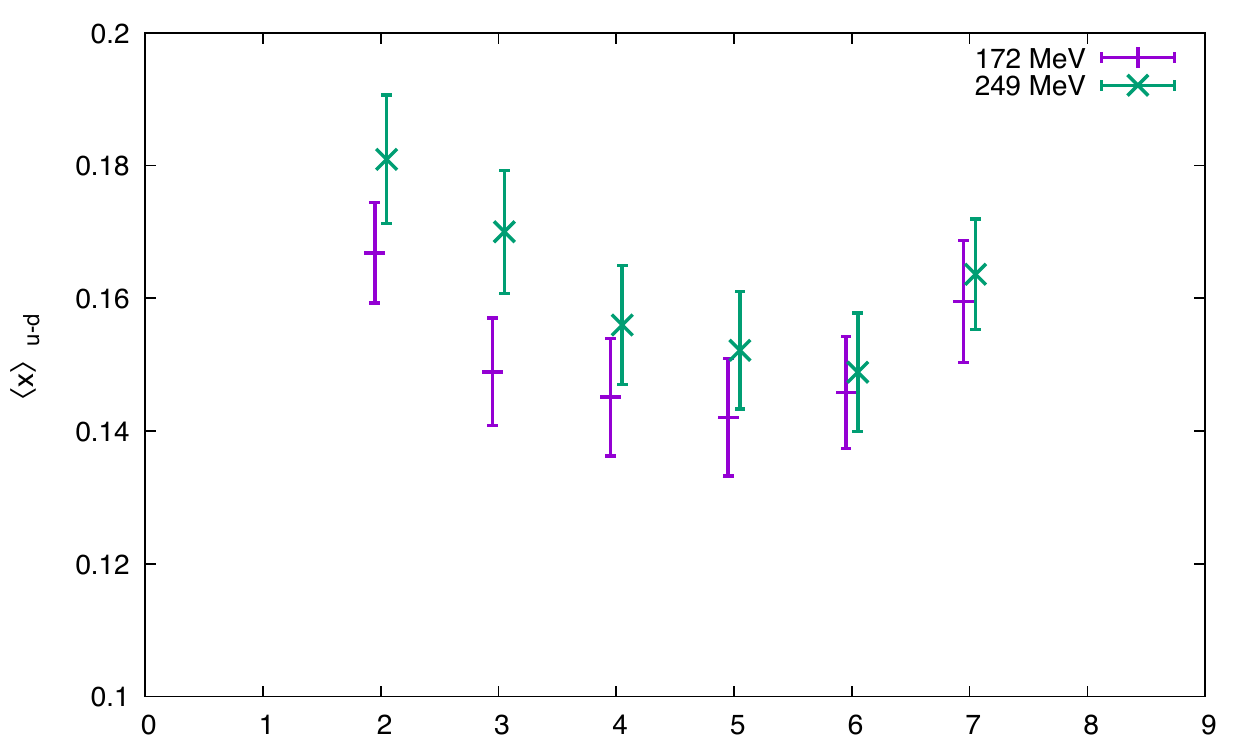}
\includegraphics[width=0.48\textwidth,clip]{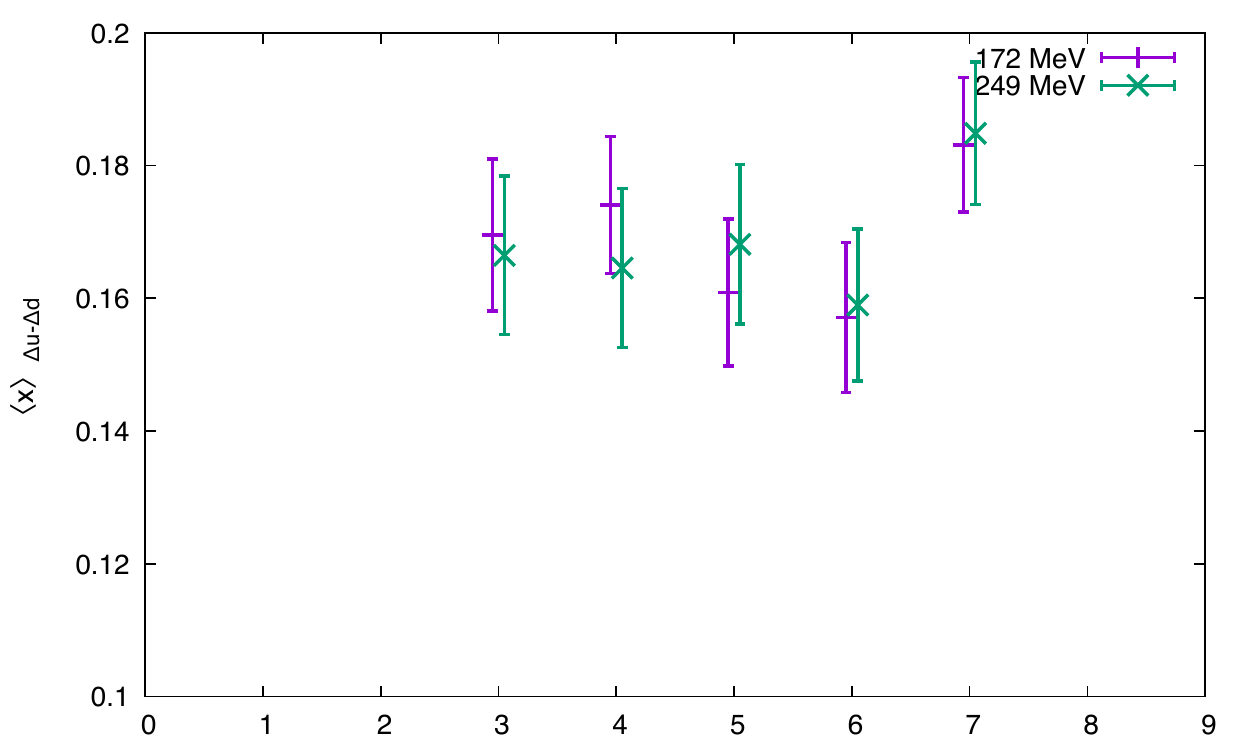}
\caption{\label{fig:fractions}
Plateau signals for the bare isovector quark momentum, \(\langle x \rangle_{u-d}\), and helicity, \(\langle x \rangle_{\Delta u-\Delta d}\).
}
\end{figure}

Signals for the isovector quark momentum, \(\langle x \rangle_{u-d}\), and helicity, \(\langle x \rangle_{\Delta u-\Delta d}\), fractions are noisier (see Fig.\ \ref{fig:fractions}.)
While the momentum fraction may still be slowly decreasing with the mass, the helicity fraction appears to stay flat.
As we are yet to renormalize these, it is not possible to compare them with their counterparts from the earlier calculations at a finer lattice spacing and heavier masses \cite{Aoki:2010xg}.
However the trending down of these observables toward the experiments seen in the earlier calculations at heavier masses has at least slowed down and possibly stopped by the present mass ranges.
Signals for the twist-3, \(d_1\), moment of the isovector polarized structure function are even noisier than the momentum and helicity fractions and are yet to provide any finite value.

\section{Conclusions}


We are finalizing these analyses and will be publishing them soon.
The ensembles were generated using four QCDOC computers of Columbia University, Ediburgh University, RIKEN-BNL Research Center (RBRC) and  USQCD collaboration at Brookhaven National Laboratory, and a Bluegene/P computer of Argonne Leadership Class Facility (ALCF) of Argonne National Laboratory provided under the INCITE Program of US DOE.
Calculations of nucleon observables were done using RIKEN Integrated Cluster of Clusters (RICC) at RIKEN, Wako, and various Teragrid and XSEDE clusters of US NSF.
I thank Michael Abramczyk, Tom Blum, Taku Izubuchi, Chulwoo Jung, Meifeng Lin, Andrew Lytle, and Eigo Shintani  for their contributions in analyzing nucleon structure reported here.
I am partially supported by Japan Society for the Promotion of Sciences, Kakenhi 
15K05064.

\bibliography{nucleon}
\end{document}